\documentclass[a4paper,11pt]{article}
\pdfoutput=1 

\usepackage{jinstpub} 
\usepackage{parskip}
\usepackage{float}
\usepackage{bold-extra}
\usepackage{amsmath}
\usepackage{amssymb}
\usepackage{subfigure}
\usepackage{times, txfonts}
\usepackage{mathtools}
\usepackage{url}
\usepackage{lineno}
\title{\boldmath CMS RPC Activities During LHC LS-$2$}

\author[z,1]{M A Shah,\note{Corresponding author.}}

\author[hh,hhhh]{,M. El Sawy}
\author[a]{A. Samalan}
\author[a]{,M. Tytgat}
\author[a]{,N. Zaganidis}
\author[b]{,G.A. Alves}
\author[b]{,F. Marujo}
\author[c]{,F. Torres Da Silva De Araujo}
\author[c]{,E.M. Da Costa}
\author[c]{,D. De Jesus Damiao}
% ...

\affiliation[a]{Ghent University, Dept. of Physics and Astronomy, Proeftuinstraat 86, B-9000 Ghent, Belgium}
\affiliation[b]{Centro Brasileiro Pesquisas Fisicas, R. Dr. Xavier Sigaud, 150 - Urca, Rio de Janeiro - RJ, 22290-180, Brazil}
\affiliation[c]{Dep. de Fisica Nuclear e Altas Energias, Instituto de Fisica, Universidade do Estado do Rio de Janeiro, Rua Sao Francisco Xavier, 524, BR - Rio de Janeiro 20559-900, RJ, Brazil}
% ...
\affiliation[hh]{The British University in Egypt (BUE), Elsherouk City,  Suez Desert Road,  Cairo 11837- P.O. Box 43,Egypt.}
% ...
\affiliation[hhhh]{Department of Physics, Faculty of Science, Beni-Suef University, Beni-Suef, Egypt}
% ...
\affiliation[z]{National Centre for Physics, Quaid-i-Azam University, Islamabad, Pakistan.}
% ...

\emailAdd{mashah@cern.ch}

\abstract{The second LHC long shutdown period (LS2) is an important opportunity for the CMS Resistive Plate Chambers (RPC) to complete their consolidation and upgrade projects. The consolidation includes detector maintenance for gas tightness, HV (high voltage), LV (low voltage) and slow control operation. All services for the RPC Phase-2 upgrade, namely RE3/1 and RE4/1, were anticipated for installation to LS2. This paper summarises the RPC system maintenance and upgrade activities.}

\keywords{Resistive-plate chambers}

\collaboration[c]{on behalf of the CMS collaboration}
\proceeding{15$^{\text{th}}$ Workshop on Resistive Plate Chambers and related detectors (RPC2020) 
\\
  Roma, Italy\\
  10-14 February 2020}
\usepackage{float}%
\usepackage{graphicx, subfigure}

\begin{document}


\maketitle

\flushbottom
\section{Introduction}
\label{sec:intro}

The CMS \cite{b} muon system, described in detail in~ \cite{d}, uses three different technologies; drift tubes (DT) in the barrel region, cathode strip chambers (CSC) in the endcap region, and resistive plate chambers (RPC) in both the barrel and endcap, and it covers a pseudorapidity region $ |\eta| < 2.4$. The DTs and RPCs in the barrel cover the region $ |\eta| < 1.2$, while the CSCs and the RPCs in the endcaps cover the eta region $0.9 < |\eta| < 2.4$. The barrel region is divided into 5 separate wheels (named $\pm 2$, $\pm 1$ and 0) while the endcaps are organised in 4 disks both in the forward and backward directions (named $\pm 4$, $\pm 3$, $\pm 2$, $\pm 1$). The $4^{th}$ RPC station has been installed during the first LHC long shutdown in 2013$-$2014 (LS1). Each wheel is divided into 12 sectors in $\phi$ while every endcap station into 36 sectors. In total there are 1056 RPC chambers, covering an area of more than 3000 m$^{2}$, equipped with 123,432 readout strips. The CMS RPCs are double-gap chambers with 2 mm gas gap width each and a copper strip readout plane located between the gas gaps. The bakelite bulk resistivity is in the range of $1-6$ $\times$ $10^{10}~\Omega$ cm and they operate in avalanche mode with a gas mixture of 95.2\% $C_2H_2F_4$, 4.5\% $iC_4H_{10}$ and 0.3\% $SF_6$ with a relative humidity of $40$-$50$\% \cite{tdr}.

In the years from 2010 to 2018 the CMS detector recorded 177.65 fb$^{-1}$ data and the RPC system contributed efficiently during the entire period. The total accumulated charge for the CMS RPC system is 2.3 mC/cm$^{2}$ for the barrel RPCs and 7.5 mC/cm$^{2}$ for the endcap RPCs.

\begin{figure}[!htb]
\centering
\includegraphics[width=.95\textwidth]{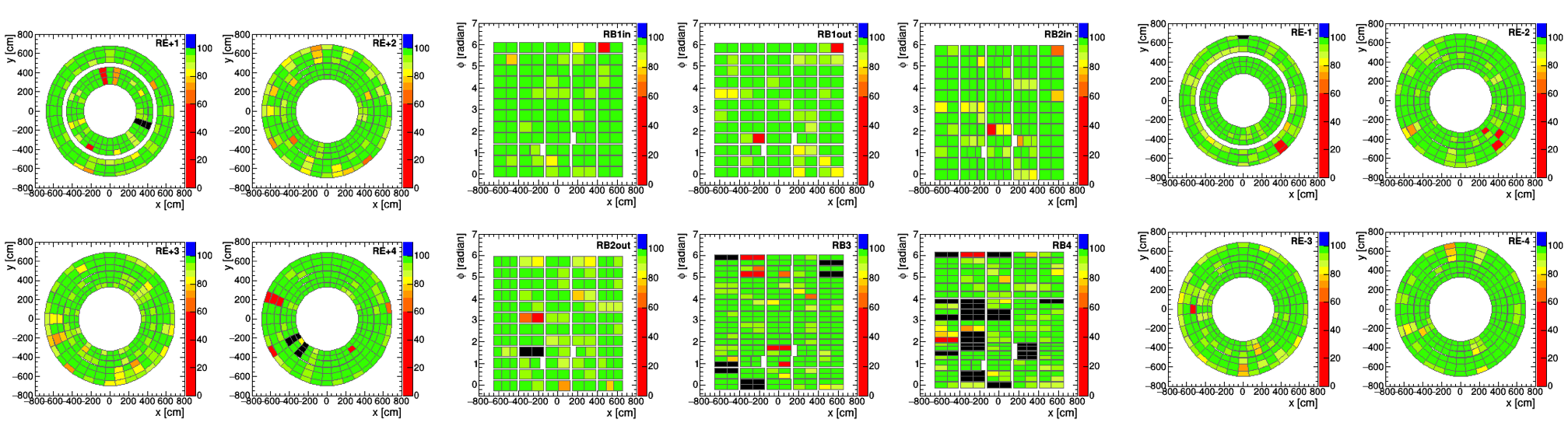}
\caption{2D efficiency maps for all barrel layers and endcap stations.}
\label{eff}
\end{figure}

Figure \ref{eff} shows the 2018 efficiency of the full RPC system (6 RPC Barrel layers and 8 Endcap stations), in longitudinal direction $z$ and azimuthal angle $\phi$ of the expected impact region of muons. Data points with low statistics or temporary problems are removed. The efficiency was obtained using the Tag-and-Probe method with a single muon triggered data set. Probe muons are reconstructed using the tracker muon algorithm, which is independent from the RPC system. They require the muon have a transverse momentum larger than 10 GeV and that at least one segment matches within a local $x$ position of 3 cm. The regions with black colour correspond to the chambers without any RPC hits, because of known hardware problems, such as gas leakage and chambers working in single gap operation mode due to HV problems. There are low efficiency or inactive regions due to spacers, chamber boundaries or masked readout strips. Full performance results with comparison to previous years are reported in Ref. \cite{rpc2018}.

The fraction of inactive electronic channels during 2018 was 3.5\%. 26 chambers were disconnected from the gas system or gas distribution racks, to reduce the leak rate. Six chambers were switched off because of HV or LV issues. 50 RPC chambers were working in single gap operation mode due to HV connector problems. 22 RPC chambers were working without properly set threshold discriminator values due to failures in the LV distribution boards. CMS is undergoing an intensive upgrade and maintenance program during the second two-year-long shutdown period (LS2). To ensure an excellent performance of the detector in the subsequent physics program, the RPC group developed in the present shutdown a thorough detector consolidation program to repair most of the hardware problems mentioned above.

\subsection{HV and LV Maintenance}
One of the important parameters to monitor when studying the RPC performance is the applied high voltage (HV). The CMS RPC achieve their optimal performance in the high voltage region [9, 9.8] kV with an average of 9.5 kV applied to each gap. The goal for the HV maintenance is to identify which part of the HV supply system is causing the current leakage and fix it in the best possible way. The LV and slow control maintenance aim at assuring proper operation and configuration of on-detector electronics, including the Front-end Boards (FEBs) and the LV distribution boards (LVDB), good functionality of the LV power boards and flawless operation along the communication bus. Around 50\% of HV and LV problems were recovered by January 2020.

\subsection{RPC Upgrade Services Installation}

The RPC Phase-II Upgrade \cite{tdr} foresees installation of all gas, cooling, and cable services for improved RPC (iRPC) during LS2. This comprises thousands of kilometres of HV \& LV cables, stainless steel gas pipes between predistribution gas racks in the Service cavern (USC) and gas distribution racks in the experimental cavern (UXC), copper pipes between distribution racks and chambers, gas impedance boxes, support equipment, and optical fibres for servicing new detectors to be installed in the near future in the innermost ring of the third and fourth endcap stations. The upgrade of the RPC gas system includes big pipework from the service through the experimental cavern and up to the CMS detector as well as significant modification of some of the existing gas racks. The cooling system for the RE4/1 detector is branched off from the existing YE3 minimanifolds while the RE3/1 chambers will be connected in series with the existing RE3 cooling loops. For optical fibers it is planned to carry out quality control tests before and after installation (in situ) using an optical time-domain reflectometer (OTDR). Installation of services for RE3/1 and RE4/1 chambers is already completed. The upgrade power system hardware, including racks, crates, power supplies, power distribution boxes, service power and communication lines, should be installed during LS2. The HV and LV power board upgrade planned for LS2 aims to replace the obsolete electronic components, and is expected to be ready for post-LS2 operations.

\subsection{RPC RE4 Activities}
\begin{figure}[!htb]
\centering
\includegraphics[width=.55\textwidth]{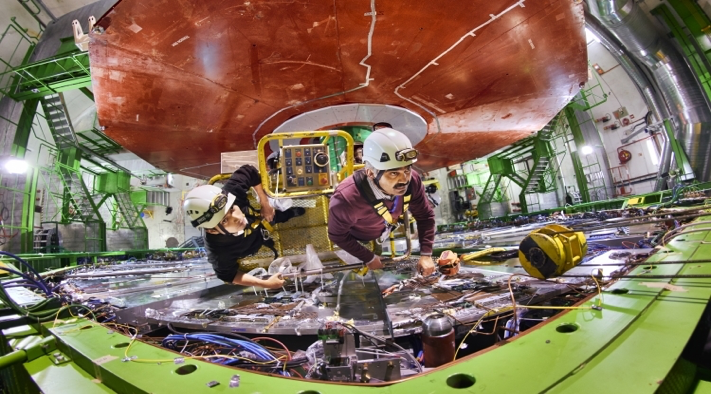}
\caption{A picture taken during the removal of one of the RE+4 supermodules.}
\label{re41}
\end{figure}

One of the key RPC LS2 interventions is dismounting 72 super modules (SM) from the fourth endcap muon stations RE4 at both endcaps to allow the CSC ME4/1 chamber extraction to replace their electronics. The first intervention on this campaign was done in March 2019 by removing 36 SM from RE+4 station. The extraction of RE4 super modules of about 4 meters long and weight of 230 kg each was challenging (see Fig. \ref{re41}). This was the very first extraction of this type of modules since RE4 station was installed during LS1.

A new lab with controlled environmental conditions, including temperature and relative humidity (T, RH), was built in an existing Point 5 building to house dismounted RE4 supermodules. A new gas line providing the standard RPC gas mixture to the surface lab was prepared by the EP-DT gas group to provide gas flow to the RE4 chambers for their commissioning. All parameters are monitored in the newly developed WebDCS RE4 interface system.

\begin{figure}[!htb]
\centering
\subfigure[]{%
\includegraphics[width=.41\textwidth]{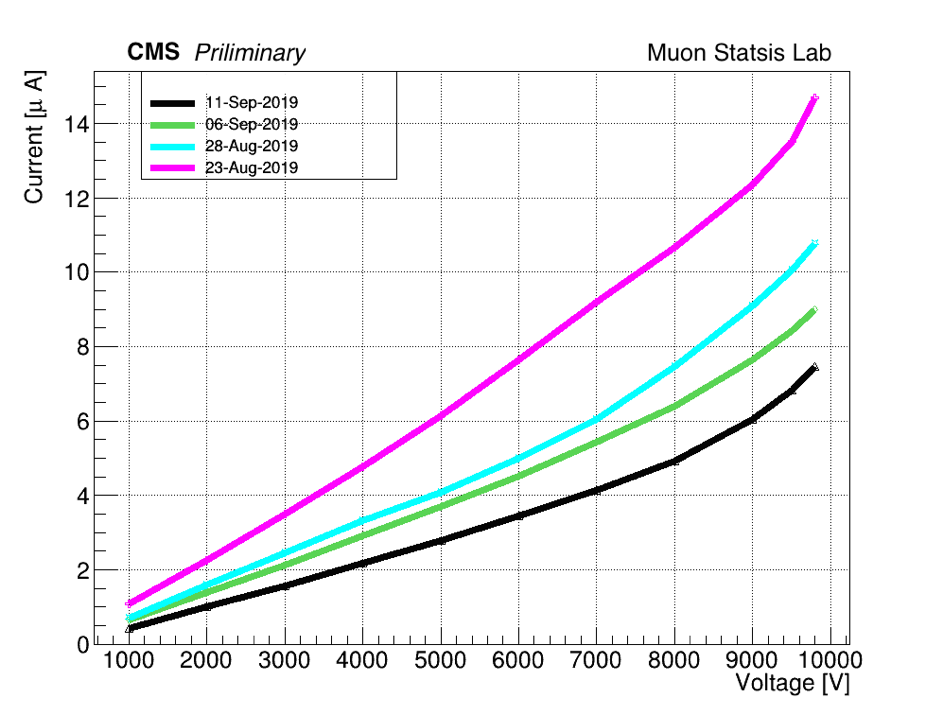}
\label{fig:subfigure1111}}
\quad
\subfigure[]{%
\includegraphics[width=.32\textwidth]{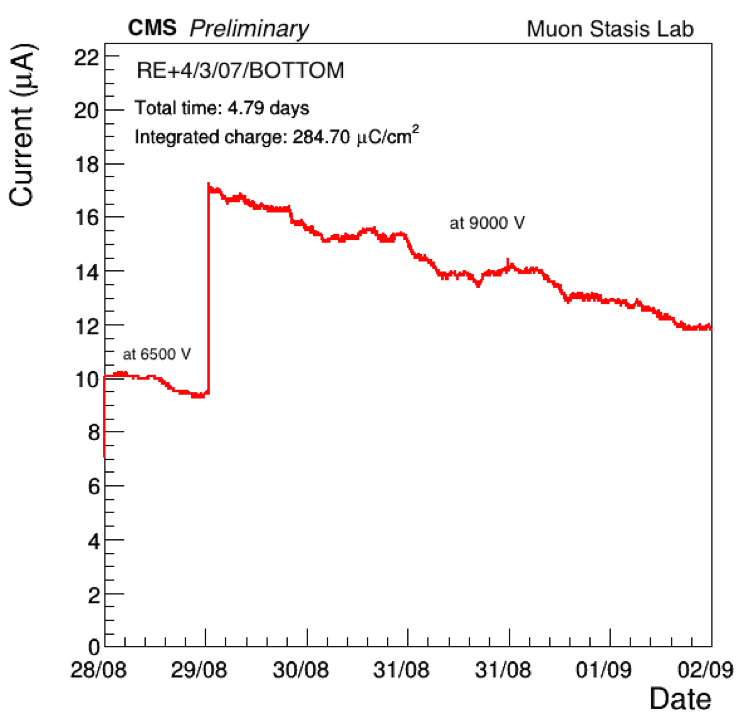}
\label{fig:subfigure222}}
\caption{Current vs HV in (a) and current vs time in (b).}
\label{re42}
\end{figure}

When powered on the surface, RE4 currents were higher than their last RUN-$2$ operational values. An example of current vs voltage dependence (VA curve) of one RE+4 supermodule commissioned in the surface lab is shown in Fig. \ref{fig:subfigure1111}. 

These chambers were in open air for several months. When chambers are kept under stable HV values for a longer periods, the currents decreased to their values at the end of RUN-$2$, as shown for one of the RE+4 chamber gaps in Fig. \ref{fig:subfigure222}. A procedure was developed and all chambers went through this recovery conditioning for four weeks. The only chamber that was not certified for reinstallation in the CMS detector because of high currents was replaced with a spare one. A possible reason for high currents could be the different environmental conditions. Studies are ongoing on the surface to understand the cause by analysing the currents with different gas humidity and recovery time of the currents with respect to the background in the regions where those chambers were installed in CMS. Dedicated noise scans were also performed to spot any dead or noisy strips. Front-end board functionality was also tested. All faulty FEBs were replaced by spare ones.

\subsection{RPC Gas System Consolidation}
Gas leaks were identified in 82 barrel RPC chambers due to cracked or broken pipes. The RPC leak repair campaign has the highest priority during LS2. 

\begin{figure}[!htb]
\centering
\includegraphics[width=.55\textwidth]{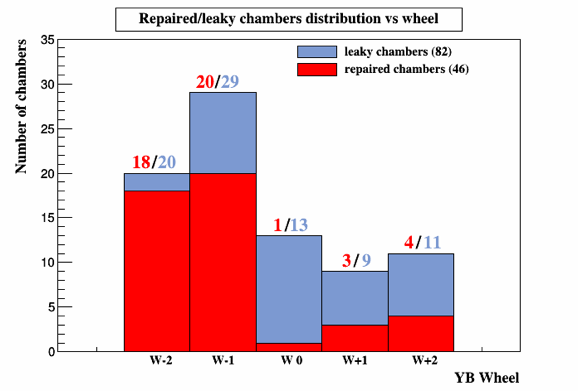}
\caption{Distribution of leaking and recovered gas channels per wheel.}
\label{leak}
\end{figure}

Dedicated tools for partial chamber extraction and milling chamber frame side C-shaped profiles were built to allow reparation of broken gas components upon identification of their exact location by means of a state-of-the-art endoscope. Activity is ongoing and Fig. \ref{leak} represents the current status of repairs with respect to the total leakage in all 5 barrel wheels.

In order to minimise pressure variations in the chambers, a possible source of new leaks, automatic pressure regulation valves on predistribution gas racks in USC are under thorough examination for forthcoming installation. The chamber pressure to be used for controlling the new automatic regulation valves will be measured by pressure sensors installed inside the new gas-tight dummy chambers. The CERN EP-DT gas group is currently carrying out an $R\&D$ program to develop the first $C_{2}H_{2}F_{4}$ recuperation system. Full details are reported in Ref. \cite{reberto}.


\section{Conclusion}
CMS RPCs have been operating very successfully during RUN-$2$. CMS is undergoing an intensive upgrade and maintenance program during the second LHC long shutdown. In order to ensure excellent detector performance in the subsequent physics program, the RPC detector experts are working hard to consolidate the RPC detector and its gas system for stable future operation. All repaired detectors and detector systems will be fully commissioned and certified for forthcoming data taking once the LHC resumes operation at the end of LS2.

\acknowledgments We would like to thank especially all our colleagues from the CMS RPC group and L1 muon trigger group for their dedicated work to keep the performance of the RPC system stable. We wish to congratulate our colleagues in the CERN accelerator departments for the excellent performance of the LHC machine. We thank the technical and administrative staff at CERN and all CMS institutes.


\end{document}